\documentclass[aps,prx,showpacs,twocolumn,amsmath,amssymb,superscriptaddress]{revtex4}
\usepackage{graphicx}
\usepackage{Jonasmacros}
\usepackage{bm}
\usepackage{mathptmx}
\usepackage{txfonts}
\usepackage{epstopdf}

\usepackage[colorlinks,citecolor=blue,linkcolor=magenta]{hyperref}
\usepackage[usenames,dvipsnames,svgnames]{xcolor}

\begin{document}
\date{\today}
\title{Magnon Dirac materials}

\author{J. Fransson}
\email{Jonas.Fransson@physics.uu.se}
\affiliation{Department of Physics and Astronomy, Uppsala University, Box 516, SE-751 21\ \ Uppsala, Sweden}

\author{A. M. Black-Schaffer}
\affiliation{Department of Physics and Astronomy, Uppsala University, Box 516, SE-751 21\ \ Uppsala, Sweden}

\author{A. V. Balatsky}
\affiliation{Nordita, Center for Quantum Materials, KTH Royal Institute of Technology, and Stockholm University, Roslagstullsbacken 23, SE-106 91\ \ Stockholm, Sweden}
\affiliation{Institute for Materials Science, Los Alamos, New Mexico 87545, USA}

\begin{abstract}
We demonstrate how a Dirac-like magnon spectrum is generated for localized magnetic moments forming a two-dimensional honeycomb lattice. The Dirac crossing point is proven to be robust against magnon-magnon interactions, as these only shift the spectrum. Local defects induce impurity resonances near the Dirac point, as well as magnon Friedel oscillations. The energy of the Dirac point is controlled by the exchange coupling, and thus a two-dimensional array of magnetic dots is an experimentally feasible realization of Dirac magnons with tunable dispersion.
\end{abstract}
\pacs{75.30.Fv, 73.22.Lp, 75.30.Hx, 75.70.Rf}
\maketitle

\section{Introduction}
\label{sec-intro}
The interest in Dirac materials (DM) as materials supporting Dirac-like spectra of excitations is rapidly growing since the discovery of graphene and topological insulators  \cite{RevModPhys.81.109,Qi_SCZhang_RMP11,RevModPhys.84.1067}. The majority of the  discussion to date has been focusing on  materials where the quasiparticles have Fermi-Dirac statistics, such as electrons. The concept of DM with a Dirac-like fermionic excitation spectrum \cite{DMreview} apply to materials ranging from graphene \cite{Novoselov_science2004,Nature.438.197} and topological insulators \cite{Kane_RMP2010}, to $d$-wave superconductors in two dimensions (2D) and the newly discovered 3D Weyl \cite{PhysRevB.83.205101,PhysRevLett.107.186806,arXiv1301.0330,Physics.4.36,AnnuRevCondensMatterPhys.5.83,PhysRevX.5.031013} and Dirac \cite{PhysRevB.85.195320,Science.343.864,PhysRevLett.113.027603} semimetals.

The concept of DM can also be applied to a wider class of quantum materials, including materials with Bose-Einstein statistics for the quasiparticle excitations. Recent advances in the synthesis of various 2D artificial structures have also spread the research on DM to artificial systems beyond bulk  materials. Examples include the theoretical prediction of artificial materials with Dirac plasmons \cite{weick2013}, photonic topological insulators \cite{khanikaev2012,tzuhsuan2015}, and superconducting grains arranged in honeycomb lattice \cite{banerjee2015}. These all point to the existence of bosonic Dirac materials (BDM). However, all these realizations of BDM  are yet to be experimentally implemented. Nevertheless, the consistency of analysis and similarities in properties of DM make it clear that this is a growing class of materials of increased scientific and technological interest.

In this article we study the properties of collective spin excitations, magnons, emerging from interacting spins in a honeycomb lattice. Considering nearest neighbor interactions as the predominant spin-spin interaction, we demonstrate that Dirac magnons emerge naturally from the spatial sublattice symmetry inherent to the honeycomb lattice. Ferromagnetic (FM) spin lattices gives Dirac-like spectra around the $K$ and $K'$ Brillouin zone corners, while the dispersion in antiferromagnetic (AFM) spin lattices is Dirac-like around the center $\Gamma$-point.
Higher order magnon-magnon interactions tend to only rigidly shift the spectrum. As the spin-spin interaction sets the scale of the magnon velocity, adapting the interaction provides direct tuning between fast and slow Dirac magnons \cite{PhysRevB.92.045401}. To illustrate the universality of DM properties across fermionic and bosonic systems we also demonstrate how local defects breaking the sublattice symmetry generate local impurity resonances \cite{DMreview}, as well as magnonic Friedel oscillations.

\begin{figure}[b]
\begin{center}
\includegraphics[width=0.99\columnwidth]{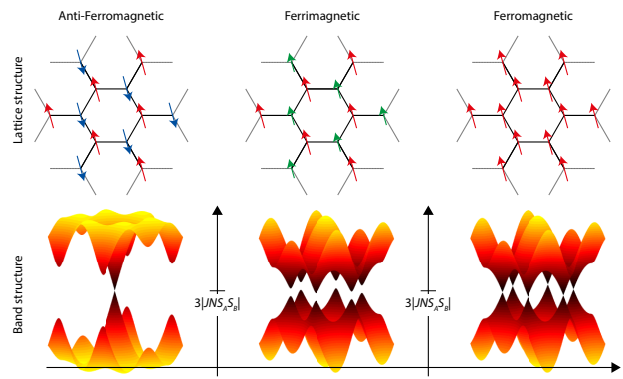}
\end{center}
\caption{(Color online)
Characteristic features of the honeycomb lattice with AFM, ferrimagnetic, and FM configurations. The upper panels show the spin lattices while the lower panels show typical calculated magnon band structures.
}
\label{fig-Fig1}
\end{figure}

Our predictions of Dirac magnons should be experimentally accessible through, e.g., engineered spin structures on metallic surfaces.
Creating 2D structures by depositing atomic or molecular magnetic absorbants on a substrate using scanning tunneling microscope (STM) provides a very natural way to create 2D honeycomb magnetic structures.
It has become routine over the past decades to use STM to engineer structures like, e.g., quantum corrals \cite{crommie1993} and artificial graphene \cite{Manoharan_Nature2012}, which makes the realization of BDM with magnons directly experimentally accessible. The surface electrons mediated magnetic interactions between the absorbant spins may range from being purely isotropic to strongly anisotropic \cite{imamura2004, fransson2010, zhou2010, khajetoorians2012}, which offer further flexibility. The unique ability of STM to freely manipulate the positions of the absorbants also allows for an almost continuous tuning of the magnetic exchange interactions which, thereby, enables full control of the Dirac magnon velocity.
The simplicity by which Dirac magnons may be generated provides an experimental advantage compared to other BDM.

The article is organized by first considering the spin wave formulation of ferro- and ferrimagnetic honeycomb lattices in Sec. \ref{sec-FM} and subsequently anti-ferromagnetic honeycomb lattices in Sec. \ref{sec-AFM}. We continue our discussion by considering the effects of local impurities, or, defects in Sec. \ref{sec-imp} and the spin wave mediated interaction, susceptibility, between pairs of impurities in Sec. \ref{sec-susc}. Thereafter, we briefly discuss the impact of higher order corrections in Sec. \ref{sec-higher} and follow up with arguments for experimental realization in Sec. \ref{sec-exp}. The paper is finally concluded and summarized in Sec. \ref{sec-conclusions}.

\section{Ferro- and ferrimagnetic lattice}
\label{sec-FM}
Our proposal of Dirac magnons can be easily justified from the effective magnon model of a general ferrimagnetic spin lattice model. In the absence of magnetic anisotropies, spin interactions are well described by the Heisenberg Hamiltonian $\Hamil_S=-\sum_{\av{ij}}J_{ij}\bfS^{(A)}_i\cdot\bfS^{(B)}_j$, where the summation runs over nearest neighbors. In a general bipartite lattice we assign spins $\bfS^{(A)}_i$ and $\bfS^{(B)}_i$ to the two different sublattices. Applying the Holstein-Primakoff transformation, assuming that the magnonic fluctuations are much smaller than the total spin $S_{A/B}=|\bfS^{(A/B)}|$, and, for now, assuming uniform ferromagnetic interactions $J_{ij}=J>0$, we can write the effective quadratic magnon model
\begin{align}
\Hamil_\text{FM}=&
	\sum_i(\dote{A}a_i^\dagger a_i+\dote{B}b_i^\dagger b_i)
\nonumber\\&
	-
	J\sqrt{S_AS_B}\sum_{\av{ij}}(a_i^\dagger b_j+{\rm H.c.})
		-3JNS_AS_B
	.
\label{eq-lattFM}
\end{align}
The first term describes the on-site magnon energies $\dote{A(B)}=3JS_{B(A)}+g\mu_BB$, where we have included an external magnetic field, while the second contribution describes the coupling between sublattices $A$ and $B$. In reciprocal space, letting $a_i=\sum_\bfk a_\bfk e^{i\bfk\cdot\bfr_i}/\sqrt{N}$ and $b_j=\sum_\bfk b_\bfk e^{i\bfk\cdot\bfr_j}/\sqrt{N}$, we obtain
\begin{align}
\Hamil_\text{FM}=&
	\sum_\bfk
	\Bigl(
		\dote{A}a_\bfk^\dagger a_\bfk
		+
		\dote{B}b_\bfk^\dagger b_\bfk
		+
		[
			\phi(\bfk)a_\bfk^\dagger b_\bfk
			+
			{\rm H.c.}
		]
	\Bigr)
\nonumber\\&
		-3JNS_AS_B
	,
\label{eq-effFM}
\end{align}
where the structure factor $\phi(\bfk)=-J\sqrt{S_AS_B}\sum_i\exp(i\bfk\cdot\bfdelta_i)$ is given in terms of the nearest neighbor vectors $\bfdelta_i$. The eigenenergies are
\begin{align}
E_\pm(\bfk)=&
	[
		\dote{A}+\dote{B}
		\pm
		\Omega(\bfk)
	]/2
	,
\label{eq-FMspectrum}
\end{align}
where $\Omega^2(\bfk)=\Delta^2+4|\phi(\bfk)|^2$, with $\Delta=\dote{A}-\dote{B}=-3J(S_A-S_B)$. Here, we then retain the normal quadratic magnon dispersion around $\Gamma$ ($\bfk=0$), see Fig. \ref{fig-Fig1} (middle and right lower panels).

\begin{figure}[t]
\begin{center}
\includegraphics[height=4cm]{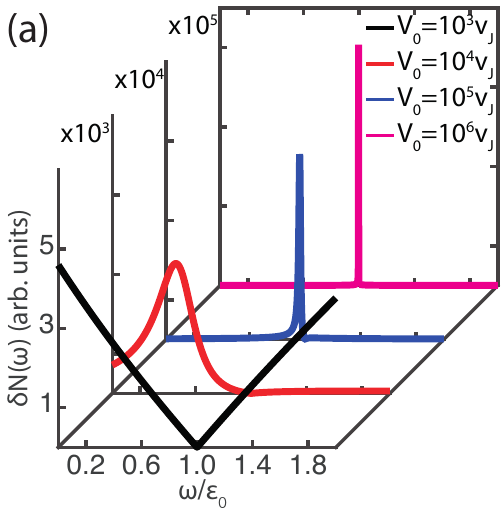}
\includegraphics[height=4cm]{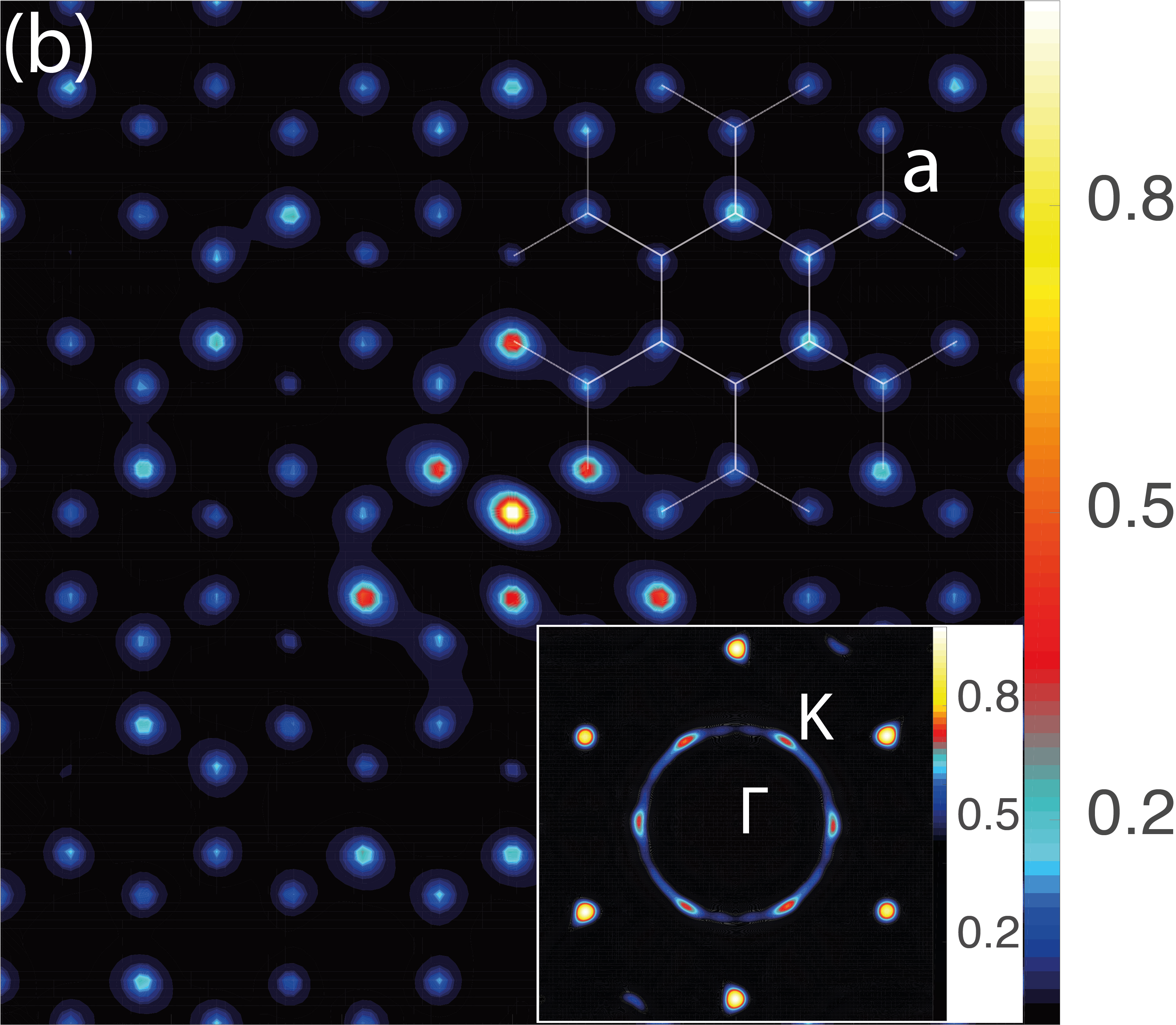}
\end{center}
\caption{(Color online)
(a) Impurity correction $\delta N(\omega)$ to the integrated density of magnon states in rising order for the scattering potential $V_0/v_J\in\{10^3,\ 10^4,\ 10^5,\ 10^6\}$.
(b) Real space distribution of the impurity correction.
Inset: Fourier transform of the impurity correction.
}
\label{fig-Fig2}
\end{figure}

For FM structures $S_{A(B)}=S$ and thus $\Delta=0$, giving $\Omega(\bfk)=|\phi(\bfk)|$. For the honeycomb lattice, where $\bfdelta_1=a(\sqrt{3},1)/2$, $\bfdelta_2=-a(\sqrt{3},-1)/2$, and $\bfdelta_3=-a(0,1)$ with lattice parameter $a$, see Fig.~\ref{fig-Fig2} this leads to the emergence of band degeneracy points at $\pm\bfK=\pm2\pi(\sqrt{3}/3,1)/3a$, around which the dispersion is linear, $\phi(\bfk\pm\bfK)\approx\pm v_Jk\exp[\pm i(\pi/3-\varphi)]$ ($v_J=3aJS/2$, $\tan\varphi=k_y/k_x$), see Fig. \ref{fig-Fig1} (right column).
In the ferrimagnetic case, on the other hand, $S_A\neq S_B$, such that $\Delta\neq0$ which leads to a gap opening at $\pm\bfK$, with the gap size $\sim3J|S_A-S_B|$ set by the difference of the spins in the two sublattices, see Fig.~\ref{fig-Fig1} (middle column).
For momenta and energies around $\bfK$, and analogously around $-\bfK$, we can thus summarize the Dirac magnon model as $\Hamil_\text{FM}=\sum_\bfk\Phi_\bfk^\dagger(\boldsymbol{\epsilon}+v_J\bfk\cdot\bfsigma)\Phi_\bfk-3JNS^2$, where the pseudo-spinor $\Phi_\bfk^\dagger=(a_\bfk^\dagger\ b_\bfk^\dagger)$ and $\boldsymbol{\epsilon}=\diag{\dote{A}\ \dote{B}}{}$. This shows that the spectrum hosts the required chirality for a DM, which directly leads to a $\pi$ Berry phase \cite{Nature.438.197,Nature.438.201} and the absence of back-scattering off smooth inhomogeneities \cite{PhysRevB.74.041403}.
These results show that a magnetic impurity honeycomb lattice provides a very simple example for realizing Dirac magnons, at the same time we cannot exclude that more complicated systems give rise to similar properties.

\section{Anti-ferromagnetic lattice}
\label{sec-AFM}
Next, we consider the emergence of Dirac bands around the $\Gamma$-point in AFM honeycomb lattices. Starting from the same Heisenberg model on the honeycomb lattice, we rotate the spins in, say, the $B$ sublattice, since the magnetism in the two sublattices point in opposite directions. Hence, for the spin operators in the $B$ sublattice we let $S_x\rightarrow-S_x$, $S_y\rightarrow S_y$, and $S_z\rightarrow-S_z$, which ensures that we capture the correct magnetic states of the $A$ and $B$ sublattices. Application of the Holstein-Primakoff and Fourier transformations lead to the quadratic effective magnon model
\begin{align}
\Hamil_\text{AFM}=&
	\sum_\bfk
	\Bigl(
		E_Aa^\dagger_\bfk a_\bfk
		+
		E_Bb^\dagger_\bfk b_\bfk
	-
		[\phi^*(\bfk)a_\bfk b_{-\bfk}
		+
		{\rm H.c.}]
	\Bigr)
\nonumber\\&
		+
		3JNS_AS_B
	,
\label{eq-effAFM}
\end{align}
where $E_A=-3JS_B-g\mu_BB$, $E_B=-3JS_A+g\mu_BB$. In this case, the magnon spectrum assumes the form
\begin{align}
E_\pm(\bfk)=&
		-\frac{1}{2}
		\Bigl(
			\Delta
			+
			2g\mu_BB
			\pm
			\sqrt{9J^2(S_A+S_B)^2-4|\phi(\bfk)|^2}
		\Bigr)
	,
\end{align}
up to the added constant in Eq.~\eqref{eq-effAFM}, making the energies overall positive. In the strictly AFM case ($S_{A/B}=S$) the energy dispersion has degeneracy points only at the $\Gamma$-point, around which we find the linear dispersion relation $E_\pm(\bfk)\approx g\mu_BB\pm\sqrt{2}v_Jk$, see Fig. \ref{fig-Fig1} (left column). While a linear dispersion near the $\Gamma$-point is expected for general AFM configured bi-partite lattices \cite{JonesMarch}, we here notice that the model can be written as $\Hamil_\text{AFM}=\sum_\bfk\Psi_\bfk^\dagger(g\mu_BB+\sqrt{2}v_J\bfk\cdot\bfsigma)\Psi_\bfk+3JNS^2$, using $\Psi_\bfk^\dagger=(a_\bfk^\dagger\ b_{-\bfk})$, which manifests the chiral Dirac nature of AFM magnons.
Finally, we note that Eq.~(\ref{eq-effAFM}) can also be obtained directly from Eq.~(\ref{eq-effFM}) by continuously changing, e.g., $S_B$ between $S_A$ and $-S_A$, manifesting the equivalence between the two models.

It should noticed that the spin wave model presented in Eq. (\ref{eq-effAFM}) is stable for magnetic fields $|B|<3J\min\{S_A,S_B\}/g\mu_B$, whereas for stronger fields collinear with the N\'eel order will generate spin flips towards the N\'eel order perpendicular to the field.

\section{Impurity scattering}
\label{sec-imp}
We now investigate the properties of impurity scattering and we, henceforth, concentrate the discussions to the FM lattices since the results are similar for AFM. Here we imagine a local defect created by removing or modifying a spin in the honeycomb lattice. We know from fermionic DM that strong local defects significantly modify the spectral function near the Dirac node \cite{DMreview}. Indeed, this is the case for the BDM as well, as we will show below. We find that a local defect induces an impurity resonance and modifies the Dirac nodal magnon spectrum. We begin by adding a single potential scattering impurity in, e.g., the $A$ sublattice, to $\Hamil_\text{FM}$. The defect is described by $\Hamil_\text{def}=V_0a^\dagger_Ia_I$, where $V_0$ is the scattering potential, and we, for simplicity, assume a ferromagnetic spin lattice. Using the $T$-matrix approach to obtain the local magnon structure we can write the single magnon Green function $\bfG_{\bfk\bfk'}(z)=\av{\inner{\Phi_\bfk}{\Phi_{\bfk'}^\dagger}}(z)$ as
$\bfG_{\bfk\bfk'}=\delta(\bfk-\bfk')\bfg_\bfk+\bfg_\bfk e^{-i\bfk\cdot\bfr_I}\calT e^{i\bfk'\cdot\bfr_I}\bfg_{\bfk'}$, where
$\calT=(\sigma^0+\sigma^z)/2(V_0^{-1}-g_0)$
whereas
\begin{align}
\bfg_\bfk(\omega)=&
	\frac{1}{(\omega-\dote{0}+i\delta)^2-|\phi(\bfk)|^2}
	\begin{pmatrix}
		\omega-\dote{0} & \phi(\bfk)
		\\
		\phi^*(\bfk) & \omega-\dote{0}
	\end{pmatrix}
\end{align}
is the bare magnon Green function. Here, $g_0(\omega)\sigma^0=\sum_\bfk\bfg_\bfk(\omega)$, $g_0(\omega)=-2(\omega-\dote{0})[2\ln(D_c/|\omega-\dote{0}|)+i\pi\sign(\omega-\dote{0})]/D_c^2$, $\dote{0}=\dote{A(B)}$, and $D_c^2$ is a high energy cut-off \cite{PhysRevB.73.125411}. We calculate the density of magnon states $N(\bfk,\omega)=-\im\bfG_{\bfk\bfk}(\omega)/\pi$, from which we obtain the sublattice resolved correction $\delta N_{A/B}(\bfk,\omega)$ given by
\begin{subequations}
\label{eq-dN}
\begin{align}
\delta N_A(\bfk,\omega)\approx&
	\frac{2}{D_c^2}
	\frac{|\omega-\dote{0}|}{|V_0^{-1}-g_0(\omega)|^2}
		\frac{(\omega-\dote{0})^2}{[(\omega-\dote{0})^2-|\phi(\bfk)|^2]^2}
	,
\label{eq-dNA}
\\
\delta N_B(\bfk,\omega)\approx&
	\frac{2}{D_c^2}
	\frac{|\omega-\dote{0}|}{|V_0^{-1}-g_0(\omega)|^2}
		\frac{|\phi(\bfk)|^2}{[(\omega-\dote{0})^2-|\phi(\bfk)|^2]^2}
	.
\label{eq-dNB}
\end{align}
\end{subequations}
The impurity scattering generates an impurity resonance within the spectrum, roughly at  $\omega=\dote{0}-D_c/2V_0$, which approaches the Dirac point in the limit of strong impurity potential, i.e. $\omega\rightarrow\dote{0}$, $V_0\rightarrow\infty$, see Fig. \ref{fig-Fig2}(a) where we plot the impurity correction to the integrated density of magnon states, $\delta N(\omega)=\sum_\bfk[\delta N_A(\bfk,\omega)+\delta N_B(\bfk,\omega)]$, for increasing $V_0$.
The impurity scattering resonance is pinned to the Dirac point for strong scattering potential, which has been shown to be a universal feature of ferminonic DM \cite{Hudson00,Gomez-Rodriguez_PRL10,Biswas10,Black-Schaffer12imp,Black-Schaffer15gapfilling, DMreview}. Our results demonstrates that this holds also for bosonic DM, not previously discussed.
It is important to point out, however, that the impurity resonance emerges only in the sublattice that does not host the defect or scattering center. Here, the defect is located in the $A$ sublattice which, in the limit of strong impurity potential, leads to the correction $\delta N_A\rightarrow0$ as $\omega\rightarrow\dote{0}$, see Eq.~(\ref{eq-dNA}), while $\delta N_B$ diverges, see Eq.~(\ref{eq-dNB}). This feature carries over to the spectral density in real space, see Fig.~\ref{fig-Fig2}(b), such that the impurity scattering induced Friedel oscillations generates an enhanced density primarily in sublattice $B$.

The scattering induced Friedel oscillations can be analyzed by evaluating the corresponding real space quantities, which are given by \cite{PhysRevB.90.241409} $\bfG(\bfr,\bfr')=\bfg(\bfR)+\bfg(\bfR_I)\calT\bfg(-\bfR_I')$, where $\bfR=\bfr-\bfr'$, $\bfR_I=\bfr-\bfr_I$, etc.. Here, $\bfg(\bfr,\omega)=g_0(\bfr,\omega)\sigma_0+\bfg_1(\bfr,\omega)\cdot\bfsigma$, where $\sigma_0$ and $\bfsigma$ are the identity matrix and vector of Pauli matrices in the pseudo-spin space, with \cite{PhysRevB.87.245404} $g_0(\bfr,\omega)=2\pi[v_Jx/iD_c^2]H_0^{(1)}(xr)\cos\bfK\cdot\bfr$ and $\bfg_1(\bfr,\omega)=2\pi[v_Jx/(iD_c)^2]H_1^{(1)}(xr)(\sin\theta\sin\bfK\cdot\bfr,i\cos\theta\cos\bfK\cdot\bfr,0)$, where $x=(\omega-\dote{0})/v_J$, $H_n^{(1)}(x)$ is the Hankel function, whereas $\theta=\phi_r+\pi/6$ with $\tan\phi_r=r_y/r_x$. The local magnon density $N(\bfr,\omega)$ around the impurity describes waves emanating from the scattering center with the wavelength set by the energy scale $(\omega-\dote{0})/v_J$ with the standard asymptotic 2D decay $\sim1/r$. The inset of Fig. \ref{fig-Fig2}(b) shows the corresponding Fourier transform with increased intensity at the ${\bf \pm K}$-points.

\section{Susceptibility}
\label{sec-susc}
Considering the susceptibility ${\cal C}(\bfr,\bfr';z)=\av{\inner{n(\bfr)}{n(\bfr')}}(z)$, where the occupation number operator $n(\bfr)=a^\dagger(\bfr)a(\bfr)+b^\dagger(\bfr)b(\bfr)$, we can make an estimate regarding its static spatial properties by approximating the $\omega\rightarrow0$ limit of the retarded form of the two-magnon propagator according to ${\cal C}^r(\bfr,\bfr')\approx-\tr\im\int n_B(\dote{})\bfG(\bfr,\bfr';\dote{})\bfG(\bfr',\bfr;\dote{})d\dote{}/\pi$, where the trace runs over pseudo-spin degrees of freedom. Partitioning of the Green function into pseudo-spin (in)dependent $\bfG_1$($G_0$) components, we can write the static susceptibility as
${\cal C}^r(\bfr,\bfr')=-2\im\int n_B(\dote{})[G_0(\bfr,\bfr';\dote{})G_0(\bfr',\bfr;\dote{})+\bfG_1(\bfr,\bfr';\dote{})\cdot\bfG_1(\bfr',\bfr;\dote{})]d\dote{}/\pi$.
From this expansion, it is easy to see for the unperturbed propagators, $\bfg$, that the susceptibility in the magnon Dirac material is given by $(\bfR=\bfr-\bfr'$)
\begin{align}
{\cal C}(\bfR)=&
	\frac{4\pi v_J^3}{D_c^4R^3}
	\sum_{n=0,1}
	\Bigl(
		1+\cos2\bfK\cdot\bfR\cos^n2\theta
	\Bigr)
	\calF_n(\dote{0})
	,
\end{align}
where
$\calF_n(\dote{0}) = \im\int[xH_n^{(1)}(x)]n_B(v_Jx/R+\dote{0})dx$, with $n_B(\omega)$ the Bose distribution function.
Here, we notice that the susceptibility is short ranged, $1/R^3$, which is in analogy with previous findings for the indirect Ruderman-Kittel-Kasuya-Yosida (RKKY) spin interaction in graphene \cite{PhysRevB.81.205416,PhysRevB.83.165425,PhysRevB.84.115119}. Although the static susceptibility ${\cal C}(\bfr,\bfr')$ given here is an average of the interactions between defects separated by $\bfR$, as it is the summation over all interactions for defects located in the same or different sublattices, its spatial characteristics is expected to have the same properties.

The magnonic mediated pair interaction between two scattering centers separated by $\bfR$ is dependent on the sublattice symmetry in a more explicit way, since defects within the same sublattice have a different pair interaction compared to defects located in different sublattices. This is in strong analogy with the RKKY interaction for magnetic defects in graphene \cite{PhysRevB.81.205416,PhysRevB.83.165425,PhysRevB.84.115119}, since the sublattice structure plays an integral role in its properties. For defects within the same sublattice, say, the $A$ sublattice, the pair interaction is provided by $\calD_A(\bfr,\bfr')=-\im\int n_B(\dote{})V_1G^{(A)}(\bfr,\bfr';\dote{})V_2G^{(A)}(\bfr',\bfr;\dote{})d\dote{}/\pi$, whereas between defects in different sublattices the interaction is given by $\calD_{AB}(\bfr,\bfr')=-\im\int n_B(\dote{})V_1G^{(AB)}(\bfr,\bfr';\dote{})V_2G^{(BA)}(\bfr',\bfr;\dote{})d\omega/\pi$. Hence, again taking the unperturbed magnon Green function and assuming that the interactions $V_n=V_0$, $n=1,2$, we find that the real parts of these interactions can be written as
$\calD_{A/B}(\bfR)=\alpha_0(1+\cos2\bfK\cdot\bfR)\calF_0(\dote{0})$, and
$\calD_{AB/BA}(\bfR)=\alpha_0[1+\cos2(\bfK\cdot\bfR\pm\theta)]\calF_1(\dote{0})$, where $\alpha_0=2\pi v_J^3V_0^2/D_c^4R^3$.
The interaction between defects is the same within the $A$ sublattice as within the $B$ sublattice, as expected due to the sublattice symmetry. For the defects in different sublattices there is a phase difference in the $A$--$B$ and $B$--$A$ interactions which is related to the conjugated phases in $\bfg_1$.
The apparent similarities between these and corresponding results for magnetic defects graphene is not surprising since both the FM Dirac magnons and graphene have the same underlying lattice structure.

\section{Higher order corrections}
\label{sec-higher}
To firmly establish the honeycomb magnon system as a BDM, we need to also consider higher order corrections. To first order the Holstein-Primakoff expansion of the spin operators leads to Eq.~(\ref{eq-effFM}), while higher order corrections provides contributions of the forms $(a^\dagger_ia_i+b^\dagger_jb_j)a^\dagger_ib_j+{\rm H.c.}$ and $a^\dagger_ia_ib^\dagger_jb_j$.
Physically, the first contribution describes hopping of magnons between sites $i$ and $j$ in sublattices $A$ and $B$, respectively, in the presence of magnons already occupying these sites, which can be regarded as magnon-assisted hopping between the sublattices in addition to the bare hopping already provided in Eq.~(\ref{eq-effFM}). The second contribution describes a direct magnon-magnon interaction for magnons in the different sublattices. Although these two contributions, as well as higher order corrections, may account for anharmonic magnon effects, we conclude that they do not change our main results, since they all preserve the symmetry established within the basic picture in Eq.~(\ref{eq-effFM}). Thus higher order effects do not destroy the bosonic DM properties. Defects of the kind discussed above, however, destroy the local sublattice symmetry, which may cause local magnon occupation imbalance between the sublattices. Under such disorder conditions, higher order contributions may provide additional effects.

\section{Experimental realization}
\label{sec-exp}
Finally, we consider a realistic and feasible route to experimentally verify our predictions. We propose that engineered spin structures on metallic surfaces provide a natural system for Dirac magnons, since this serves as a standard way to create 2D magnetic structures by depositing magnetic defects on a substrate using STM. 
It has been established that surface electron mediated magnetic interactions between the deposited spins may range all from being purely isotropic to strongly anisotropic \cite{imamura2004,fransson2010,zhou2010,khajetoorians2012}.
Here, since the materials specific details are not important, except that the surface should have a metallic band of surface states, we adopt a simple continuum model for a metallic 2D electron gas onto which the localized spins $\bfS_i$ are adsorbed
\begin{align}
\Hamil=&
	\sum_{\bfk\sigma}
		\psi_\bfk^\dagger\boldsymbol{\epsilon}(\bfk)\psi_\bfk
		-
		g\mu_BB\sum_iS_{iz}
		-
		\sum_{\bfk\bfk'i}v\bfs_{\bfk\bfk'}\cdot\bfS_i
		.
\end{align}
The first term describes the surfaces states of the substrate, in terms of the spinor $\psi_\bfk=(\cs{\bfk\up}\ \cs{\bfk\down})^T$, given some dispersion $\boldsymbol{\epsilon}(\bfk)$. Hence, for a non-magnetic metallic surface we may use a diagonal quadratic dispersion, while spin orbit (SO) coupling can introduce chirality into the model. An external magnetic field $\bfB=B\hat{\bf z}$ is also applied to the system, by which we can control the effective chemical potential for the spin system. The spin moments $\bfS_i$ interact locally via exchange ($v$) with the electron spins $\bfs_{\bfk\bfk'}=\psi_\bfk^\dagger\bfsigma\psi_{\bfk'}$ in the surface. By distributing the spin moments in a regular lattice, we can refer to this model as a Kondo lattice.

For metallic surface states and sufficiently large separation of the spins, it is well-known that the effective spin-spin interaction is mediated by the surface electrons, as described by the RKKY interaction \cite{ruderman1954,kasuya1956,yosida1957}. This interaction can generally be partitioned into an isotropic Heisenberg, $J_{ij}$, and anisotropic Ising, $\mathbb{I}_{ij}$, and Dzyaloshinski-Moriya, $\bfD_{ij}$, interactions \cite{imamura2004,fransson2010,fransson2014}, providing the effective spin Hamiltonian
\begin{align}
\Hamil_S=&
	-\sum_{ij}\bfS_i\cdot(J_{ij}\bfS_j+\bfD_{ij}\times\bfS_j+\mathbb{I}_{ij}\cdot\bfS_j)
	.
\end{align}
Here, e.g., for metallic surfaces without any magnetic texture the Heisenberg interaction parameter $J_{ij}=-2v^2\im\int f(\omega)\calG_{ij}^r(\omega)\calG_{ji}^r(\omega)d\omega/\pi$, where $\calG_{ij}^r(\omega)$ is the retarded real space Green function for the surface states between $\bfr_i$ and $\bfr_j$, whereas $f(\omega)$ is the Fermi distribution function. For large separation $R_{ij}$ between the spins $J_{ij}\sim\cos(2k_FR_{ij})/(2k_FR_{ij})^2$, where $k_F$ is the Fermi wave vector, such that we can obtain FM ($J_{ij}>0$) or AFM ($J_{ij}<0$) exchange interaction depending on the spatial separation. Furthermore, the quadratic spatial decay of $J_{ij}$ and $\mathbb{I}_{ij}$ \cite{fransson2010} suggests that a nearest neighbor interaction between the spins will describe the physics sufficiently well for metals with small or negligible spin-orbit coupling. Hence, our effective spin model reduces to nearest neighbor interactions. For metals with strong spin-orbit coupling, a non-negligible chiral next-nearest interaction from the linearly decaying Dzyaloshinski-Moriya interaction is introduced in the effective model. This can be detrimental to the Dirac magnons, since it can introduce parity breaking contributions to the magnon model, Eq. (\ref{eq-effFM}), which breaks the sublattice symmetry.
Counteracting this parity breaking is the appearance of a finite Ising interaction dyad $\mathbb{I}_{ij}=I_{ij}\hat{\bf z}\hat{\bf z}$ for spin-polarized metals. This term serves as a stabilizing mechanism for the spin moments along some quantization axis. Hence, a collinear ground state can in fact still be favored, with the effective Hamiltonian for the magnons is given by either Eq.~(\ref{eq-effFM}) or (\ref{eq-effAFM}).

\section{Summary and conclusions}
\label{sec-conclusions}
Beginning from an isotropic Heisenberg model of localized spin moments in a honeycomb lattice, we address the possibility of emerging Dirac magnons. Our finding suggest that the magnonic band structure for the ferromagnetic lattice closely resembles the fermionic band structure for, e.g., graphene. In this sense, there are Dirac point degeneracies at the $K$ and $K'$ points in the Brillouin zone around which the energy-momentum dispersion relation can be linearized. The Dirac point occurs at an elevated energy as the magnon spectrum is defined for positive energies only, i.e., there is no Fermi level around which the band structure is centered. The band structure changes upon variations in the magnetic ordering and we find a gap opening at the Dirac point for ferrimagnetic lattices, with the size of the gap scaling with the difference of the spins in the two sublattices according to $3J|S_A-S_B|$. Pushing the magnetic ordering to a fully anti-ferromagnetic lattice redistributes the band structure such that there is only one Dirac point occurring at the $\Gamma$-point. In analogy with fermionic Dirac materials, we find impurity resonance near, and pinned to, the Dirac point for local impurities/defects in the lattice structure, suggesting universality of impurity induced resonances. Likewise is the susceptibility and pair interaction between local defects in the lattice governed by similar spatial properties as for graphene. We have, furthermore, shown that the Dirac spectrum is robust agains higher order contributions in the spin wave expansion of the Heisenberg model. Finally, we propose that the type of our suggested magnonic Dirac material should be possible to engineer by distributing magnetic adatoms in a honeycomb lattice on metallic surface. Choosing metallic substrate lacking magnetic texture, e.g., Cu surface should guarantee that anisotropic Ising and Dzyaloshinski-Moriya interactions can be avoided.

\acknowledgments
We are grateful to G.~Aeppli, Z.~Huang, and S.~Ming for useful discussions. This work was supported by the Swedish Research Council, the G\"oran Gustafsson Foundation, the Knut and Alice Wallenberg Foundation, the Wallenberg Academy Fellows program, the Swedish Foundation for Strategic Research, and the European Research Council under the European Union's Seventh Framework Program (FP/2207-2013)/ERC Grant Agreement DM-321031. Work at Los Alamos was supported by the US DoE BES E304.
\bibliography{DiracMaterials}

\end{document}